\documentclass{emulateapj}
\usepackage{apjfonts}
\usepackage{epsfig,graphicx}

\slugcomment{Received 22 January 2007; Accepted 20 February 2007}

\shorttitle{The Milky Way: An Exceptionally Quiet Galaxy}

\shortauthors{Hammer et al.}

\begin{document}

\title{The Milky Way: An Exceptionally Quiet Galaxy;\\
Implications for the formation of spiral galaxies
}

\author{F. Hammer, M. Puech, L. Chemin, H. Flores, and M. D. Lehnert}
\affil{GEPI, Observatoire de Paris, CNRS, Univ. Paris 7, \\
5 place Jules Janssen, 92195 Meudon France}
       \email{francois.hammer@obspm.fr}

\begin{abstract}
The Milky Way has been generally considered to be representative
of the numerous spiral galaxies inhabiting the local Universe, thus
providing general and perhaps the most detailed constraints on numerical
models of galaxy formation.  We compare both the Milky Way and M31
galaxies to local external disk galaxies within the same mass range,
using their locations in the planes drawn by $V_{flat}$ versus $M_{K}$
(the ``Tully-Fisher'' relation), $j_{disk}$ (angular momentum) and 
the average Fe abundance, [Fe/H],  of stars in the outskirts of the galaxy.
These relations are thought to be the imprints of the dynamical,
star-formation, and accretion history of disk galaxies.  We compare
the best established Tully-Fisher relations and reconcile their slopes
and zero points in the plane $M_{K}$-$V_{flat}$. We then compare the
properties of local spirals from a representative sample to those of
the Milky Way and M31 considering how these two galaxies would appear
if observed at larger distances.

We find, for all relationships, that the Milky Way is systematically
offset by $\sim$ 1$\sigma$ or more from the distribution of comparable
local galaxies: specifically, it shows a too small stellar mass,
angular momentum, disk radius and outskirts stars [Fe/H] ratio at
a given $V_{flat}$, the latter being taken as a proxy for the total
mass. In contrast with the Milky Way, M31 lies well within the mean of
the fundamental relationships. On the basis of their locations in the
($M_{K}$, $V_{flat}$ and $R_{d}$) volume, the fraction of spirals like
the Milky Way is 7$\pm$1\%, while M31 appears to be a ``typical'' spiral.
As with M31, the bulk of local spirals show evidence for a formation
history shaped mainly by merging. The Milky Way appears to have had an
exceptionally quiet formation history and had escaped any significant
merger over the last $\sim$10 Gyrs which may explain why its angular
momentum, stellar mass and [Fe/H](outskirts) are two to three times
smaller than those of other local spirals. We conclude that the standard
scenario of secular evolution driven by the accretion of gas and disk
instabilities is generally unable to reproduce the properties of most
(if not all) spiral galaxies, which are well represented by M31. However,
the relatively recent proposal explaining the evolution of spiral galaxies
through merging (the so-called ``spiral rebuilding'' scenario of Hammer
et al. 2005) is consistent with the properties of both distant galaxies
(e.g., stellar mass assembly through episodic IR luminous burst phases
driven predominately by mergers) as well as to those of their descendants
-- the local spirals.

\end{abstract}


\section{Introduction}

Disk galaxies constitute the majority of the galaxy population observed in
the local universe. They represent 70\% of intermediate mass galaxies
(stellar masses ranging from 3 $\times$ 10$^{10}$ to 3 $\times$
10$^{11}$ M$_{\odot}$), which themselves include at least two-third of
the present-day stellar mass \citep[e.g.,][]{Hammer05}. Early studies
of the Milky Way have led to a general description of the formation of
a disk galaxy embedded in a halo \citep{ELS62}. \citet{Fall80} set out
a model of galaxy formation assuming that disks form from gas cooling
and condensing in dark halos. Protogalactic disks are assumed to be
made of gas containing substantial amount of angular momentum, which
condenses into stars to form thin disks \citep{Larson76}.  These disks
then evolve only through secular processes. This so-called standard
model successfully reproduces the flat rotation curves and the size of
spirals \citep[e.g.,][ and reference therein]{Mo98}. Such a model is
(still) particularly adept at reproducing the properties of the Milky
Way \citep[e.g., ][ and references therein]{Naab06}.

However, there are several outstanding difficulties with this
standard scenario. One such difficulty is the so-called angular
momentum problem.  That is, simulated galaxies cannot reproduce
the large angular momentum observed in nearby spiral galaxies
\citep[e.g.,][]{Steinmetz99}. Another is the assumed absence of
collisions during and after the gas condensation process. Indeed, the
hierarchical nature of the $\Lambda$CDM cosmology predicts that galaxies
have assembled a significant fraction of their masses through collisions
with other galaxies.  It is likely that such collisions would easily
destroy galactic disks \cite[e.g.,][]{Toth92}. Although the accretion of
satellites may preserve the disk, it is also true that major collisions
would certainly affect it dramatically.  The key questions are then:
Do major collisions always destroy disks? Can major collisions lead
to the formation of new disks?  Do these rebuilt or altered disks have
properties consistent with those of local galaxies?

Observations of the merger rate evolution have now reached sufficient
maturity to provide useful constraints on theory of galaxy evolution. For
example, considering only galaxies with mass larger than 3 $\times$
10$^{10}$ M$_{\odot}$, the pair fraction of galaxies over the relative
mass range of 1:1 to 1:3 at z=0.6 is $\sim$ 5$\pm$1\% \citep[see][ and
references therein]{Bell06}. There is a remarkable agreement between
different methods of estimating the pair fraction.  All recent estimates,
no matter what the technique, for example, two point correlation
techniques \citep{Bell06}, or pair counts \citep{LeFevre00},
or morphological classifiers (CAS: \citealt{Conselice03}, GINI:
\citealt{Lotz06}) give consistent results.  However, to constrain
the cosmological evolution of the merger rate requires us to assume a
characteristic time for a real pair to actually merge.  Using arguments
based on either dynamical friction \citep{Binney87} or simple orbital
time-scale \citep[e.g.,][]{Bell06}, this time scale has been estimated to
be about 0.35 Gyrs.  Combining the pair fraction and characteristic time
scale estimates suggests that for a present-day galaxy with a stellar
mass larger than 3 $\times$ 10$^{10}$ $M_{\odot}$, the chance it has
experienced a major merger since z=1 is  50$\pm$17\%, 75$\pm$25\% and
70\% according to \citet{Lotz06}, \citet{Hammer05}, and \citet{Bell06},
respectively\footnote{The differences, although small, are probably
related to disagreements about the slope of the pair fraction redshift
evolution, possibly due to different approaches in correcting for the
effects of evolution or not.}.  Although less certain, integrating the
merger rate to higher redshift implies that a typical bright galaxy
may have experienced up to four to five major merging events since z=3
\citep{Conselice03}.

The high frequency of major mergers may be a real problem for the
standard theory of disk formation.  Assuming that protogalactic disks
lie in the distant universe, how can this be reconciled with an absence
of major collisions?  How can we explain the large fraction of local
disks if major mergers (with mass ratio ranging from 1:1 to 1:3)
unavoidably lead to the formation of an elliptical?  Even at z$\le$1
the observations are challenging for the standard scenario. At least one
third of intermediate mass galaxies at z$=$ 0.4-1 have morphologies very
discrepant from that of E/S0/Sp \citep{Brinchmann98, vandenBergh01,
Zheng05}. A similar fraction of distant galaxies possess complex
velocity fields \citep[26\%,][]{Flores06, Puech06} Peculiar morphology,
and even more so, complex kinematics are almost certainly a result of
on-going or recent mergers \citep{Puech06, Puech07}. If those galaxies
were the progenitors of present-day ellipticals, this would lead to a
much larger fraction of ellipticals than what is observed. Less than
10\% of local intermediate mass galaxies are ellipticals \citep[7\%
of galaxies brighter than $M_{B}$ $<$ $-$20;][]{Conselice06}, and they
have formed the bulk of their stellar mass earlier, likely before z=1
\citep[see][]{Jimenez06, Bernardi06}.

In summary, observations of distant galaxies pose a challenge to the
standard secular scenario of spiral formation and evolution.  The past
history of the Milky Way is certainly lacking any major (and maybe even
significant minor) interaction over at least the last 10 Gyrs. The
validity of the standard scenario is highly dependent on whether
or not the Milky Way is representative of the general population of
spiral galaxies.  In the following, we consider several probes of
the past history of galaxies and compare the properties of the Milky Way and
M31 to those of other spirals selected from a complete sample of nearby
galaxies. For
such a purpose, we cannot use optical spectral energy distribution or
colors, which can be seriously affected by instantaneous star formation
and extinction. To derive unambiguously the age and the metallicity
requires studies of individual stars. Such analyses are therefore limited
to only two massive galaxies, the Milky Way and M31.  

Galaxy dynamics (disk velocity and angular momentum) is certainly an
interesting avenue to test the history of galactic disks. The
relationship between flat rotation velocity and stellar mass
is very tight \citep[e.g.,][]{Verheijen01} and reveals the way the stellar
mass has been assembled into galactic halos \citep[see][]{McGaugh05}. The
disk angular momentum (as the product of the disk radius to the
rotational velocity) is a relic of events (or absence of events) which
have been experienced by a galaxy. It is noteworthy that the standard
scenario of disk formation shows some difficulties in reproducing both
relationships (Tully-Fisher and $j_{disk}$-$V_{flat}$). On the basis of
three dynamically related parameters (e.g., $V_{flat}$, $M_{K}$, used as
a proxy of the stellar mass and $R_{d}$), one may identify whether the
Milky Way (and M31) is representative or not of the general population
of spiral galaxies.

Few studies have brought the question of the representativeness
of the Milky Way to the attention of the astrophysical community.
\citet{Flynn06} showed that the Milky Way lies at 1 $\sigma$ from the
Tully-Fisher relations of \citet{Verheijen01} and \citet{Pizagno05}. This
conclusion needs to be firmly established. First because the two
relations taken as references show different slopes and zero points,
and second because I-band luminosity may be not accurate enough to
robustly estimate the stellar mass. More surprising is the fact that
the stellar content of the outskirts\footnote{We use content of the
{\it outskirts} to mean the stellar component that partially fills the
dark halo from 5 to 30 kpc from the galaxy centre \citep[see][and the
discussion in \S 5]{Brown06}.} of the Galaxy is apparently different (in
its stellar chemical abundances and colors) from other spiral galaxies
\citep{Mouhcine06, Zibetti04}. In the following, we attempt to establish
a Tully-Fisher relation for local spirals that is consistent for all data
sets, discuss the representativeness of M31 and the MW compared to local
spiral galaxies, and as a results of these discussions, try to understand
which scenario may apply to the entire ensemble of spiral galaxies.

The paper is organized as follows: in \S 2, we describe the properties
of the Milky Way and M31 as if they were observed at larger distances;
in \S 3, we concentrate on establishing an homogenized Tully-Fisher
relation in ($M_{K}$-$V_{flat}$) which reconciles results from
the SDSS \citep{Pizagno06} to those from \citet{Courteau97} and
\citet{Verheijen01}; in \S 4, we estimate how the Milky Way and M31
are representative of local spiral galaxies, on the basis of their
positions in the ($M_{K}$, $V_{flat}$ and $R_{d}$) volume; in \S~5,
we discuss the general evidence that the Milky Way has had a ``quiet''
merging history; and finally, in \S 6, we compare the relative merits of
the two disk formation scenarios (the standard scenario in which disk
evolution is driven mainly by secular processes and another in which
disk evolution is mainly driven by mergers of galaxies) in reproducing
the characteristics of spiral galaxies. In this paper we adopt the
Concordance cosmological parameters of $H_0$=70 km s$^{-1}$ Mpc$^{-1}$,
$\Omega_M=0.3$ and $\Omega_\Lambda=0.7$.

\section{Milky Way and M31 properties for comparison to those of other
field spirals}

\subsection{Disk scale length $R_{d}$ and angular momentum}

Measurements of the Galactic disk scale length have lead to heterogeneous
results since the early 90s. The difficulty in making such estimates
probably results from the fact that we
lie within it. Studies of the local stellar kinematics by Hipparcos
and determinations of the dust and stellar mass distributions from
all sky surveys \citep[especially at IR wavelengths; see e.g., ][ and
references therein]{Drimmel01} have greatly improved the estimates
of the Galactic disk scale length.  \citet{Sackett97} convincingly
showed that for most studies, the determination of $R_{0}$/$R_{d}$
(where $R_{0}$ is the distance to the Galactic Center) is more secure
observationally than that of $R_{d}$. Table 1 summarises the results
for 15 estimates made since 1990, assuming the most accurate and direct
estimate of $R_{0}$ \citep[i.e., $R_{0}$= 7.62$\pm$0.32 kpc,][see also
Appendix A]{Eisenhauer05}.  As Table 1 shows, very different approaches
in estimating $R_{d}$ produce a remarkably narrow range of values. Thus
it now appears reasonable to derive a value for the Galactic disk scale
length with an accuracy roughly similar to those derived for external
galaxies. One study has however produced a very discrepant result.
In this study, \citet{Mendez98} have used low-latitude star-counts from
the Guide Star Catalog to derive a value of $R_{d}$ that is roughly
twice that of other studies. While it is beyond the scope of this paper
to argue for or against the robustness of any of these results, we note
that it is quite surprising that shallow IR sky surveys such as COBE,
DIRBE, 2MASS or DENIS (see Table 1) would find systematically low values
of $R_{d}$ in comparison. Given this and the relatively large quoted
uncertainty in the study of \citeauthor{Mendez98}, we choose to consider
the other 14 studies of Table 1 for our analysis.  Even if we did include
this study and take a weighted average of the ensemble of values, we
note that it would make very little difference in the resulting value.
In order to compare the value of $R_{d}$ for the MW with measurements
made in the optical (generally R or I band) for external galaxies,
we adopt $R_{d}$=2.3$\pm$0.6 kpc for the Milky Way.  This value is
within the 1$\sigma$ uncertainty of 14 of the 15 estimates presented
in Table 1. Because of possible systematic errors in we have adopted an
uncertainty in the value of $R_{d}$ that is twice the dispersion in the
14 estimates listed in Table 1.  Because \citet{Reyle01} determined
a similar value for the exponential scale length of the thick disk
(2.5 kpc), our adopted value applies for all the components with a disk
geometry in the Milky Way and is probably close to what would be derived
by an observer located outside the Milky Way.

For M31, we find that $R_{d}$=5.9$\pm$0.3 kpc in R band (or 6.5$\pm$0.3
kpc in B band), after correcting the original value in \citet{Walterbos87}
to a distance of 785 kpc \citep[see][]{McConnachie05}. \citet{Geehan06}
derive $R_{d}$=5.4 kpc using R band data from \citet{Walterbos87} and
using the same supplemental data from \citet{Kent83}. Using Spitzer
observations, \citet{Barmby06} find $R_{d}$=6.08$\pm$0.09 kpc. These
values are unlikely to be affected by the outer, star-forming ring
of M31 which is weak at red wavelengths. Moreover, Barmby et al.
carefully estimated the disk scale length excluding areas containing
the outer ring and the most distant regions of the galaxy that are
likely to be relatively more contaminated by emission from the sky
background. Furthermore, the Spitzer data reach much lower relative
surface brightness levels and thus extend much farther out in the
galaxy light profile compare to the 2MASS data \citep{Seigar06}. It
is also likely that the latter may be affected by sky substraction
\citep[see][]{Barmby06, Seigar06}.  Given this situation, we adopt
$R_{d}$=5.8$\pm$0.4 kpc, where the adopted uncertainty accounts for the
range of all the above estimates.

\begin{table}
\centering
\caption{Recent estimates of the Milky Way exponential disk scale-length\label{table_sclgth}}
\begin{tabular}{c|c|cc|cc}
\hline\hline
Ref.\tablenotemark{a}    & data set\tablenotemark{b} & $R_{0}$/$R_{d}$  & $\Delta$($R_{0}$/$R_{d}$) & $R_{d}$\tablenotemark{c} &
$\Delta$($R_{d}$)\tablenotemark{d} \\
&  &    &   & (kpc) & (kpc) \\
\hline
1   &  2.4 $\mu$m   & 2.7    &  -      &  2.8    &  0.5  \\
2   &  0.45-0.55 $\mu$m   & 3.4    &  -      &  2.23   &  0.3  \\
3   &  kinematics  & 3.1    &  0.4    &  2.5    &  0.8  \\
4   &  0.36-0.65 $\mu$m   & 3.5    &  -      &  2.17   &  0.6  \\
5   &  IRAS+NIR	   & 3      &  -      &  2.5    & -	\\
6   &  DENIS	   & 3.7    &  -      &  2.05   & 0.1	\\
7   &  2MASS	   & 4.0    &  -      &  1.9    & 0.3	\\
8   &  DIRBE	   & 2.86   &  -      &  2.66   & -	\\
9   &  kinematics  & 3.22   &  0.2    &  2.36   & 0.1	\\
10  &  kinematics  & 4.54   &  0.5    &  1.67   & 0.2	\\
11  &  M dwarves   & 2.8    &  -      &  2.7    &  0.4  \\
12  &  Guide Stars & 1.42   &  0.4    &  5.36   & 2	\\
13  &  visual	   & 3.2    &  -      &  2.38   &  0.5  \\
14  &  COBE/IRAS   & 3.55   &  -      &  2.14   &  0.1  \\
15  &  COBE/DIRBE   & 3.57   &  -      &  2.13   &  0.1  \\
\hline
\end{tabular}

\tablenotetext{a}{Notes on columms:\\
1: Kent et al (1991);
2: Robin et al (1992); 3: Fux \& Martinet (1994);
4: Ojha et al (1996); 5: Ortiz \& Lepine (1993);
6: Ruphy et al (1996); 7: Porcel et al(1998);
8: Spergel et al (1996);
9: Dehnen \& Binney (1998); 10: Bienaym\'e (1999);
11: Gould et al (1996); 12: Mendez \& van Altena (1998);
13: Ng et al (1996); 14: Chen et al (1999);
15: Drimmel \& Spergel (2001)}
\tablenotetext{b}{The data sets used to derive the values of $R_{d}$.  If the data
are from optical or near-IR imaging, we list the central wavelengths
used for the determination.} 
\tablenotetext{c}{All values are derived using a solar radius
R0=7.62 $\pm$ 0.32 kpc (Eisenhauer et al 2005).}  
\tablenotetext{d}{The quoted uncertainties are the 
the maximum of the uncertainty quoted by the
reference and of the uncertainty associated to the relation $R_{d}$=$R_{0}$
$\times$ $(R_{0}/R_{d})^{-1}$. Notice that for two studies, 5 and 8,
uncertainties are not available.}
\end{table}

We adopt $V_{flat}$= 220 km s$^{-1}$ for the Milky Way \citep[the current
IAU standard,][]{Kerr86} and 226 km s$^{-1}$ for M31 \citep{Carignan06},
adopting a conservative uncertainty of $\pm$10 km s$^{-1}$ in each value
(see a detailed discussion in Appendix A). With these values, the angular
momentum of M31 is 2.5 times higher than that of the Milky Way. The
Milky Way has indeed a small disk scale length: at $V_{flat}$= 220 km
s$^{-1}$, galaxies in the SDSS show an average disk scale length of 4.75
kpc \citep[][see their Figure~20]{Pizagno06}, twice the Milky Way value.
M31 has a disk scale length rather similar to the average value for SDSS
galaxies. These comparisons are based on what are presently the best
estimates of the disk scale length and velocity for both the Milky Way
and M31.

We are however cognisant of the fact that because these estimates have
shown some variations in the past, future experiments (such as GAIA)
will provide us with much more accurate, and perhaps even discrepant
values. However given the concordance of previously determined values,
this seems unlikely.  Perhaps more problematic therefore is the fact
that the disk scale length for the Milky Way has been estimated using
methodologies different from that used for external galaxies, including
M31. Some systematic uncertainty might affect these estimates when
comparing them with external galaxies such as M31. To make a definite
conclusion about the robustness of these estimates certainly requires a
careful analysis of the complexity of both the Milky Way and M31, which
is unfortunately beyond the scope of this paper. But we note that both
the Milky Way and M31 have been fully imaged by COBE, DIRBE, 2MASS and
DENIS for the former, and including Spitzer for the latter. Thus any
possible systematic effects related to the extinction are unlikely to
be a significant source of bias affecting the estimates of the ratio of
the disk scale lengths of MW and M31.  The robustness of the estimates
for the Milky Way are also supported by the excellent agreement between
IR measurements and models constrained by the detailed kinematics of
various galactic components (see Table 1).  Critically, both types
of methods provide estimates which should be representative of the
underlying stellar mass distribution.  The Spitzer observations are
also the best representation of the stellar mass distribution of M31
\citep{Barmby06}. Little doubt should therefore be left that M31 disk
has a significantly larger scale length that that of the Milky Way after
comparing \citeauthor{Barmby06} value for M31 ($R_{d}$=6.08$\pm$0.09 kpc)
to that all sky surveys in IR for the Milky Way ($R_{d}$=2.31$\pm$0.36
kpc, see Table 1).  We then assume in the following that the disk scale
length of M31 is 2.5$\pm$0.8 times that of the Milky Way.  We note that
the quoted uncertainty of 0.8 also takes into account possible systematic
uncertaintes in the disk scale lengths of the MW and M31.

\subsection{Total absolute luminosity in K band and total stellar mass}

Motivated by a strong desire to remove the imprint of the strong Milky
Way signal from the very faint CMB fluctuations, the COBE experiment has
provided an accurate value for the near IR luminosity of the Milky Way.
\citet{Drimmel01} derived an extinction corrected value for the K-band
absolute magnitude of $M_{K}$= -24.02. This estimate is close to $M_{K}$=
-24.12 found by \citet{Kent91}. Observations by Spitzer of M31 give
m(3.6$\mu$m)=-0.34 after sky substraction and integrating the light
profile to large distances \citep{Barmby06}. \citeauthor{Barmby06}
assumes K-m(3.6$\mu$m)= 0.3, which implies $M_{K}$=-24.51. This value
has not being corrected for extinction, the correction is $-$0.188 mag
after applying the formalism of \citet{Tully98}. The values of the K-band
absolute magnitude for both the Milky Way and M31 are robust and we adopt
a conservative $\pm$0.1 magnitude for the uncertainty in each estimate.

One is able to derive stellar mass from $M_{K}$, using an
empirical estimate of the color dependent $M_{star}/L_{K}$ ratios
\citep{Bell03}. Using B-V=0.79 \citep{Boissier00} and B-R=1.5
\citep{Walterbos87} for the Milky Way and M31, respectively, we derive
$M_{stellar}$= 5 $\times$ $10^{10}$ $M_{\odot}$ for the Milky Way and 10.3
$\times$ $10^{10}$ $M_{\odot}$ for M31. We have assumed  $M_{K\odot}$=
3.3 \citep{Bell03} and a Kroupa IMF. For the Milky Way, our estimate
is remarkably consistent with that of \citet{Flynn06}. For M31, our
estimate is very close to the sum of the disk (7.2 $\times$ $10^{10}$
$M_{\odot}$) and of the bulge (3.2 $\times$ $10^{10}$ $M_{\odot}$) found
by \citet{Geehan06}, and consistent with that of \cite[][ and Barmby,
private communication]{Barmby06}.  Because stellar mass estimates are
subject to systematic uncertainties related to the choice of the IMF
and the star-formation history through the adopted value of $M/L_{K}$,
we choose in the following, to use $M_{K}$ values as the foundation for
making the comparison between the Milky Way, M31, and other local spirals.
$M_{K}$ will be used as a surrogate for the total stellar mass.

\section{Towards an homogenised Tully-Fisher relation for local spirals}

Our goal is to derive a Tully-Fisher relation for a representative sample
of local galaxies. Measurements of a sample of SDSS galaxies ($H\alpha$
emission line) have been recently presented by  \citet[][hereafter
called the SDSS sample]{Pizagno06}.  Although more needs to be done on
this type of sample, the current sample includes all type of spirals
from S0 to Sd, and the only morphological pre-selection is requirement
that the galaxies be roughly edge-on (b/a $<$ 0.6). The present data
is apparently representative of the local galaxy luminosity function
for M$_{r}<-$20.5 galaxies, and the completeness of this sample as it
relates to our analysis will be addressed later in this section \citep[see
Figure~1 of][]{Pizagno06}.

Among the best studied Tully-Fisher relation is the study made
by \citet{Verheijen01} based on a sample of Ursa Major cluster
galaxies. \citet{Verheijen01} was able to calculate $V_{flat}$
for 28 galaxies (hereafter called the Ursa Major sample) among 38
cluster members with HI data. Verheijen mentioned that $V_{flat}$
can be estimated for all galaxies except those with rising velocity
curves. Interestingly, such sources are flagged 3 in the \citet{Pizagno06}
study, and so for further comparison between the two samples, we will
only keep flag 1 and 2 galaxies in the \citeauthor{Pizagno06} sample
\footnote{\citet{Pizagno06} classify galaxies into several distinct
catagories based on their rotation curves.  Galaxies catagorized
as flag-1 are those with extended flat portions in velocity with
increasing radius; flag-2 are those with rotation curves just reaching
the turnover region, flag-3 have rotation curves still rising at the
outermost measurement, while galaxies catagorized as flag-4 have
velocity curves that are not characterizable as rotation curves}.
Indeed $V_{flat}$ has generally been preferred to $V_{max}$, because
conversely to the latter, it is not affected by the influence on
the dynamics due to the bulge. \citet{Verheijen01} found a very tight
correlation between $M_{K}$ and $V_{flat}$ and concluded that $V_{flat}$
is the best proxy for the total galaxy mass.

Using different samples often leads to different slopes and zero
points of the Tully-Fisher relation. This is illustrated by Figure~16
of \citet{Flynn06} in which the individual Tully-Fisher relations
from \citet{Bell01} and \citet{Pizagno05} are compared. Note however
that \citeauthor{Bell01} sample is originally the \citet{Verheijen01}
sample of galaxies for which stellar masses have been additionally
estimated. Because SDSS and Ursa Major studies have applied the same
procedure to estimate extinction corrections \citep[the ``mass
dependent extinction'' method, see][]{Tully98}, we have been motivated to
understand what are the causes of this difference.

In a panel of Fig.~\ref{comp_Verheijen}, we compare the K' magnitudes
from \citet{Tully96} and 2MASS for galaxies in Ursa Major.  It shows
an excellent agreement between the measurements, except for objects
with K' $>$ 10 for which K' data overestimate the K band luminosity by
approximately 1 magnitude. Indeed, this problem was already noticed by
\citet{Tully96}, who writes, ``From inspection of the luminosity profiles,
it is seen that there is a good agreement between the various pass-bands
except that the K' material is truncated about 2 magnitudes shallower than
the B, R, I material. It would require long exposures to reach surface
brightness at K' comparable to those at optical bands''.  Because in
the 2MASS data base, magnitude errors from 0.02 to 0.08 are given, we
have decided to adopt K magnitudes from 2MASS and then compare again
the SDSS and \citet{Verheijen01} Tully-Fisher relations.

To do so, also requires a common method for estimating the rotation
velocity of galaxies within the two samples. For this, we adopt the definition of
$V_{flat}$ as a good proxy for the total mass, because:

\begin{figure}
\plotone{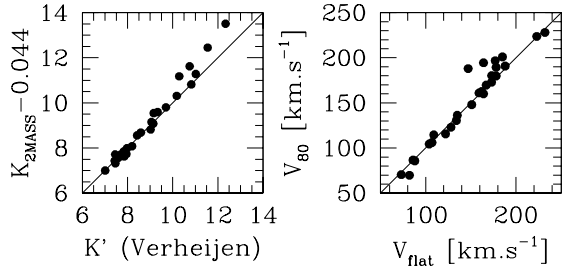}
\caption{{\it Left:} comparison between K band magnitudes from 2MASS
\citep[corrected by 0.044 mag, see][]{Bessel05} to the K' value adopted
by \citet{Verheijen01}. Although they show a very good agreement for
bright galaxies, the depth of the early measurements by \citet{Tully96}
over-estimate the luminosity of faint objects by approximately 1
magnitude. {\it Right:} Comparison of $V_{80}$ (rotational velocity
at 80\% of the total I-band luminosity) from \citet{Pizagno06} with
$V_{flat}$ from \citet{Verheijen01}. Values are very similar for all
objects except for 5 which are highly discrepant. The most discrepant
object is NGC4138 which is a Sa galaxy with $V_{80}$=187 km s$^{-1}$
and $V_{flat}$= 147 km s$^{-1}$.  NGC4138 has a large bulge and so
it is likely that $V_{80}$ is closer to $V_{max}$ than is $V_{flat}$.
Other similarly discrepant objects show large inclinations in excess
of 80$^{\circ}$.
\label{comp_Verheijen}}
\end{figure}

\begin{itemize}
\item of the remarkably tight baryonic Tully-Fisher relation over five
decades in baryonic mass ($\sim$ $V_{flat}^{4}$, see McGaugh, 2005).
Such a tight relation must indicate that it is fundamentally a correlation between the maximum
rotational velocity of the dark matter ($V_{flat}$) and the total baryonic
mass inside that halo \citep[e.g.,][]{Verheijen01};

\item it is derived at large galacto-centric radii and is thus not
affected by possible non-axisymmetric gas motions due to the presence
of a bulge or a bar, as is often the case for the definition of $V_{max}$;

\item the Milky Way and M31 have similar total masses\footnote{with
M=7.5$^{+2.5}_{-1.3} \times$10$^{11}$ M$_{\odot}$ for M31 (on the basis of
the giant stream kinematics: \citealt{Ibata04}; \citealt{Geehan06}; on the
basis of the satellite motions: \citealt{Evans00}) and  M=5 to 9 $\times$
10$^{11}$ M$_{\odot}$ \citep{Battaglia06}, for the dark matter halo of
the Milky Way, assuming either a Navarro Frenk \& White or a Truncated
Flat model, respectively.} \citep[see][and references therein]{Ibata05}
and similar $V_{flat}$ but different $V_{max}$.  Please keep in mind however
that the methods used to derive the total masses for both objects are
different.

\end{itemize}

The right panel of Fig.~\ref{comp_Verheijen} shows that estimates of
$V_{80}$ provide a good estimate for $V_{flat}$, at least for objects
without large bulges and which are not purely edge-on. Indeed, the
estimates of $V_{80}$ have been adopted by \citet{Pizagno06} to derive the
Tully-Fisher relation, as it generally samples the rotational velocity
at the optical radius and is approximately equivalent to three times
the disk scale length.

\begin{figure}
\plotone{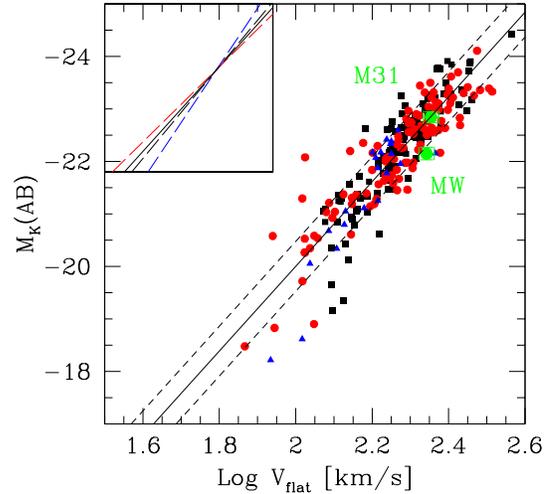}
\caption{K band Tully-Fisher relation for three local galaxy samples:
(blue) triangles \citep[22 points,][]{Verheijen01}, (red) dots
\citep[105 points,][]{Pizagno06} and (black)  squares \citep[124
points,][]{Courteau97}. After revising K band magnitudes for the
Verheijen's sample (see Fig.~\ref{comp_Verheijen}), the three samples
define  remarkably consistent relationships. The solid line is the fit
of the three samples and the dashed lines represent the $\pm$1$\sigma$
deviation about the best-fit line (0.47 mag).  This result is very
close to what has been derived from the single \citeauthor{Pizagno06}
sample. The location of the Milky Way and M31 are marked as large green
dots. In the insert at the top-left, we show the best-fit of the relation
for each of the three samples (blue dashed line: Ursa Major; black-dashed
line: UGC and red-dashed line: SDSS), illustrating the larger slope for
the Ursa Major sample (see text and Table 2).
\label{TF_all}}
\end{figure}

Fig.~\ref{TF_all} shows the Tully-Fisher relation for the two samples
\citep[][assuming the K magnitudes from 2MASS]{Verheijen01} and
\citet[][using only flag 1 and 2 galaxies]{Pizagno06}. We have also superimposed
the data from \citet[][hereafter called the UGC sample]{Courteau97},
which includes 169 Sb-Sc UGC galaxies for which 2MASS photometry
are available. For consistency, we consider only those galaxies
having rotational velocities rising with radius by less than 10\% of
the adopted $V_{flat}$, which has been assumed to be the rotational
velocity at three times the disk scale length \citep[e.g., similar to
the $V_{80}$ of][see Fig.~\ref{comp_Verheijen}]{Pizagno06}. For all
galaxies, to determine the absolute K-band magnitude, $M_{K}$(AB) (where
$M_{K}$(AB)=$M_{K}$(Vega)+1.87), we adopt the K band magnitude from 2MASS
and a single scheme for estimating the extinction \citep[using the mass
dependent extinction method][]{Tully98}.  In addition, all K-band absolute
magnitudes have been k-corrected by $-2.1\times z$ \citep{Bell03}.
The results  are given in Table 2 for each of the three samples.

\begin{table*}
\begin{center}
\caption{Slopes, zero points and residual sigmas (in magnitude)
of the fitted Tully-Fisher relations for the Ursa Major, SDSS, and UGC samples.}
\label{tab2}
{\scriptsize
\begin{tabular}{c|c|c|c|c||c|c|c|c}
\hline\hline
Sample   & $N_{\rm points}$ & Zero point $\pm 1\sigma$ error & Slope $\pm 1\sigma$ & $\sigma_{\rm res}$ & $N_{\rm points}$ &Zero point $\pm 1\sigma$ error & Slope $\pm 1\sigma$ & $\sigma_{\rm res}$ \\
\hline
UMaj 	 & 22   &  2.37        2.17    & -10.86       0.99      &  0.48     & 14  & -13.86        4.47    &  -3.68         1.99     &  0.35     \\
SDSS     & 105  & -5.90        0.78    & -7.16        0.34      &  0.47     & 79  & -6.55         1.33    &  -6.88         0.57     &  0.38     \\
UGC      & 124  & -2.29        0.93    & -8.77        0.41      &  0.44     & 97  & -4.29         1.23    &  -7.92         0.53     &  0.36     \\
All      & 251  & -3.86        0.58    & -8.07        0.26      &  0.47     & 190 & -5.96         0.87    &  -7.17         0.37     &  0.38     \\
SDSS/UGC & 229  & -4.50        0.60    & -7.79        0.26      &  0.47     & 176 & -5.60         0.91    &  -7.32         0.39     &  0.38     \\	 
\hline\hline
\end{tabular}
}
\tablecomments{In columns 2, 3, 4, and 5, we give the best fit parameters
including all galaxies for which the rotation curve is of sufficient
quality or has the appropriate characteristics necessary to derive an
accurate estimate of $V_{flat}$ (see text for details). In the columns
6 through 9, we give the best fit relations only considering galaxies
with log(V$_{\rm flat}$) $>$ 2.2. Notice that the SDSS and UGC samples
show very similar best fit parameters, and that limiting the sample
to galaxies with high rotation velocities does not alter the best fit
Tully-Fisher relations for these two samples.}
\end{center}
\end{table*}

\begin{figure}
\plotone{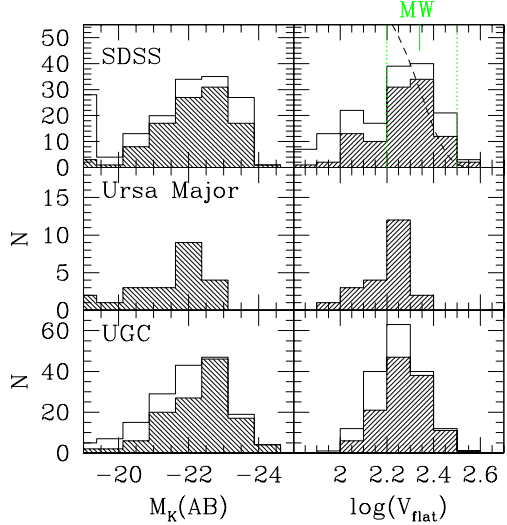}
\caption{Histograms of M$_{K}$(AB) ({\it left}) and log (V$_{\rm flat}$)
({\it right}) for the three samples, SDSS, Ursa Major and UGC ({\it top to
bottom} respectively). The shaded
histogram represents galaxies for which $V_{80}$ appears to be an appropriate
proxy for $V_{flat}$
(see text for details). The
top-left panel is very similar to the Figure~1 of \citeauthor{Pizagno06}
and shows a deficiency of faint, slowly rotating galaxies with  M$_{r}$
$>$ -21, M$_{K}$(AB) $>$ -22 and log(V$_{80}$)$<$2.2. This is illustrated
by the dashed line (top-right panel) showing the local log($V$) number
distribution function derived from the local K-band luminosity function
(Cole et al., 2001 with $M_{K}^{\star}$(AB) = $-$22.4 and a slope equal
to $-$1)
after assuming (log(V)$^{\star}$= 2.35 from the Tully-Fisher relation
and L$_{K}$  $\sim$ V$^{4}$ following \citet{McGaugh05}.  Within 2.2 $<$
log(V$_{flat}$) $<$ 2.5 (vertical dotted lines), the SDSS sample provides
a robust representation of the spiral galaxy distribution for comparison
with the Milky Way.
\label{histo3samples}}
\end{figure}

After our homogenization of data for the three samples, we find very
good agreement between UGC and SDSS data, while the Ursa Major sample
still shows a discrepant Tully-Fisher relation (Fig.~\ref{TF_all}
and Table 2). Even if the latter sample is much smaller than
the other ones, this might present a significant problem for our
purpose here.  Fig.~\ref{histo3samples} shows the distribution of
$M_{K}$(AB) and $log(V_{flat}$) for the three samples. Examination of
Fig.~\ref{histo3samples} (and of Fig.~\ref{TF_all}) is illuminating:
the Ursa Major sample includes fainter and slower rotating galaxies than
the other samples, and these galaxies tend to lie off of the relation
defined by the brighter galaxies.  This is probably due to low mass
galaxies having large gas fractions \citep[e.g.,][]{McGaugh05}. This
effect explains in a simple way the higher slope of the relation for
Ursa Major. It is also illustrative to compare the K-band luminosity
function of the Tully-Fisher samples to that of the population of
local galaxies. The SDSS sample is indeed lacking small galaxies
(see Fig.~\ref{histo3samples}), especially below log($V_{80}$)$=$2.2
(or equivalently with $M_{K}$(AB) $>$ $-$22).  Given the selection
procedure adopted by \citet{Pizagno06}, as well as the large and
statistically robust sample of the SDSS itself, we believe that it
provides the best way to test the representativeness of the Milky Way
among galaxies having parameters in the same range. The Milky Way has
log($V_{flat}$)=2.34 and Fig.~\ref{histo3samples} shows that the SDSS
sample is limited to log($V_{flat}$)=2.5, because very few spirals have
rotational velocities in excess of 320 km s$^{-1}$. Within the range
of 2.2 $<$ log($V_{flat}$) $<$ 2.5, the distribution of velocities in
the SDSS sample matches reasonably well what one would expect from a
Schechter function. In the following, we have chosen this interval in
which to characterize the representativeness of the Milky Way and M31
compared to the ensemble of local spiral galaxies.

\section{Representativeness of the Milky Way and of M31 in the ($M_{K}$,
$V_{flat}$ and $R_{d}$) volume }

We confirm the results of \citet{Flynn06} and find that the Milky Way
lies at $\sim$1$\sigma$ from the Tully-Fisher relation derived from
the three local samples (Fig.~\ref{TF_all}). Such a discrepancy for the
Milky Way is found in both the I and K band Tully-Fisher relations, so
it is unlikely that it could be affected by an error in the magnitude
or extinction estimate. \citet{Flynn06} have extensively discussed the
possible source of systematic errors associated with such estimates,
and found none convincing (see also Appendix A). The K band measurement
by COBE \citep{Drimmel01} is certainly as accurate as K band measurements
of external galaxies and accounts for extinction as well as for spiral
arms on the opposite side of the galaxy from the Sun. The Milky Way K-band
luminosity is half the average value for local spirals with similar
$V_{flat}$. This can be translated into a similar factor in stellar
mass. The difference ($\sim$ 0.7 mag) is much larger than the uncertainty
in the total K-band luminosity estimate of the Milky Way. Besides this,
M31 lies on the average relation delineated by other local spirals.

\begin{figure}
\plotone{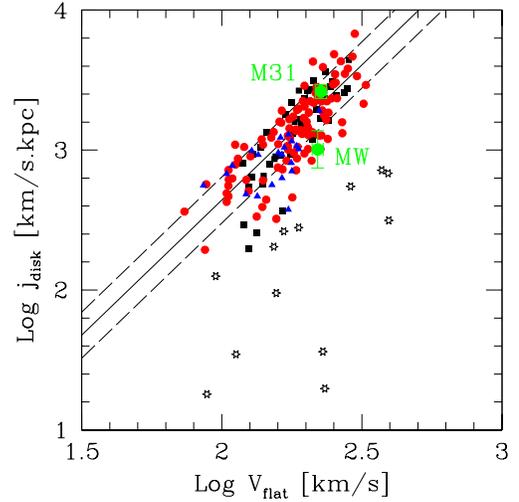}
\caption{ $j_{disk}$ versus V$_{flat}$ for galaxies in the three local
samples (same symbols as in Fig.~\ref{TF_all}). Open stars represent
the results from simulations of \citet{Steinmetz99}.  The differences
between the simulations and the data illustrate the
difficulty of the standard model for disk formation and
evolutoin in reproducing the angular momentum of spiral galaxies.
\label{j_v}}
\end{figure}

We also compute the specific angular momentum of the disk, estimating
it using  $j_{disk}$=2 $R_{d}$$V_{flat}$, appropriate for a thin disk
\citep[see][]{Mo98}. The only difference adopted here is the use of
$V_{flat}$ instead of $V_{max}$, as we believe it is a better proxy
for the halo velocity (see previous section). Disk scale lengths are
estimated in I band for the SDSS sample, K band for the Ursa major
sample, and r band for the UGC sample. We choose not to apply for an
inclination correction to $R_{d}$ values, such \citet{Dutton06} have
done, simply because we do not find any correlation between $R_{d}$
and disk inclination.  Fig.~\ref{j_v} shows that the three samples show
a similar distribution in the $j_{disk}$-$V_{flat}$ plane, and that
observed galaxies show a larger angular momentum than expectationed based
on the simulations of \citet{Steinmetz99}.  It also illustrates that,
due to its small disk radius, the Milky Way is deficient in angular
momentum by a factor 2 compared to average local spiral with the same
velocity. Conversely, M31 lies marginally above, but still well in the
mean relation of local spirals.

\begin{figure}
\plotone{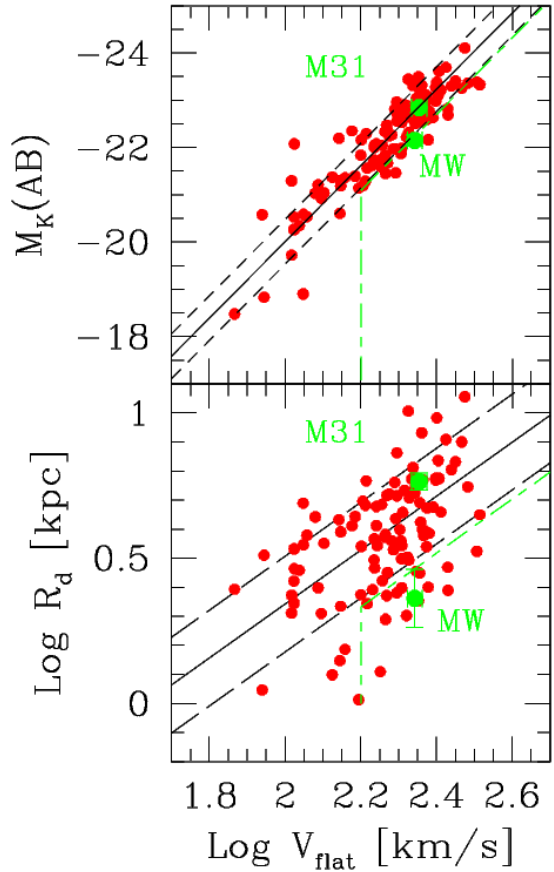}
\caption{M$_{K}$ (above) and R$_{d}$ (bottom) versus V$_{flat}$ for
SDSS galaxies (red points). Large green dots represent the position of
the Milky Way and the 1$\sigma$ uncertainty of both relations is shown
as black dashed lines. Dotted green lines show how we select Milky Way
like galaxies, which are descrepant in both $L_{K}$ and disk length
scale. Because of the large scatter in the $R_{d}$-$V_{flat}$ relation,
we also consider as an alternative estimate, all objects within $V_{flat}$=2.2-2.5
having $R_{d}$ $\le$ 2.3+0.6 kpc, in guaging the representativeness of the
Milky Way (see text for details).
\label{selection}}
\end{figure}

Assuming that the SDSS sample is a good representation of local
galaxies with 2.2 $<$ log($V_{flat}$) $<$ 2.5, one can estimate how
representative the Milky Way and M31 are in the ($M_{K}$, $V_{flat}$
and $R_{d}$) volume. Fig.~\ref{selection} illustrates how we identified
those galaxies having a deficiency in $L_{K}$ and $R_{d}$, like the
Milky Way does. We find that the fraction of Milky Way-like spirals is
7$\pm$1\% in this volume. M31 falls just on the Tully-Fisher relation
delineated by the SDSS sample (as well as by the UGC sample). However
its radius (5.8 kpc) is slightly larger than the average (4.7 kpc) at
$V_{flat}$=226 km s$^{-1}$. Using the same method as we use to determine
the Milky Way's representativeness, we find that 30$\pm$3\% of the SDSS
galaxies have a disk radius larger or equal to 5.8-0.4 kpc, i.e., those
being M31-like galaxies. Because the $R_{d}$-$V_{flat}$ distribution is
not as tight as the Tully-Fisher relation (see Fig.~\ref{selection}),
we adopt two different schemes for estimating the fraction of MW-like
(and M31-like) galaxies in the corresponding plane. One is assuming that
the $R_{d}$-$V_{flat}$ correlation is real, the other is considering
only galaxies with disk radii  smaller than 2.3+0.6 kpc (for Milky Way
like galaxies) and larger than 5.8-0.4 kpc (for M31-like galaxies). Both
alternatives produce similar numbers and the difference is attributed
to the uncertainties discussed previously.

In summary, we find that very few galaxies have ($M_{K}$, $V_{flat}$ and
$R_{d}$) properties similar to that of the Milky Way, while M31 is far
more representative. The above calculation is affected by the fact that
we have assumed a much larger relative error bar ($\Delta$$R_{d}$/$R_{d}$)
for the Milky Way than for M31. Applying similar relative errors for both
objects (i.e., considering $\Delta$$R_{d}$/$R_{d}$=0.26), would lead to
a much higher fraction of M31-like galaxies. This is consistent with the
location of the two galaxies relative to the 1$\sigma$ error of the
two relations shown in  Fig.~\ref{selection}. If the distributions
were gaussian and independent, the location of the Milky Way outside the
1$\sigma$ error is consistent with few percent of Milky Way-like galaxies,
while M31 is a typical spiral.

\section{The almost untouched outskirts of the Milky Way}

We have shown that dynamical properties of the Milky Way may be
quite exceptional compared to a local spiral galaxies with similar
rotation speeds. Let us investigate if previous kinematic events have
left some imprint on the outskirts of the Galaxy's stellar populations since
those populations are most likely to show the most obvious residual
effects of the merging history \citep[dynamical relaxation times are
long and so mixing of the stellar populations both in terms of chemistry
and dynamically is likely to be significantly less than in the disk or
inner regions of the galaxy, e.g.,][]{Font06, Renda05}. The definition
of the stellar halo by \citet[][see also \citealt{Chapman06}]{Renda05}
is intended to include all stars within the outskirts of a galaxy
(they used limiting radii ranging from 4 to 30 kpc). This definition is
comparable to the ``spheroid'' of M31 which is used by \citet{Brown06}
to be 5 to 30 kpc  from the disk minor axis. The ``spheroid'' of M31 is
well described by a de Vaucouleurs law \citep{Pritchett94, Irwin05}
up to 30 kpc \citep{Durrell04}.  Given this, we have adopted the word
``outskirts'' to encompass the various, perhaps ill-defined definitions
of what constitutes stars beyond the relatively high surface brightness
disks of these spiral galaxies.

One of the most spectacular events currently under investigation in the
outskirts of the Milky Way is the Sagittarius stream \citep{Ibata94,
Majewski03}. While it is interesting to study the Sagittarius for
understanding the evolution of the halo of the MW, its stellar content is very small
\citep[2 $\times$ 10$^{7}$ L$_{\odot}$ in V band, see][]{Majewski03}.
The Sagittarius stream represents only a very tiny fraction of the
stellar content  of the Milky Way stellar halo \citep[2 $\times$
10$^{9}M_{\odot}$, see][]{Carney90}. The stellar content of the outskirts
of the Milky Way is essentially made of old stars with low metal abundance
([$<$[Fe/H]$>$=$-$1.6, \citealt{Beers95, Morrison03}, see also the review
by \citealt{Prantzos06}). Conversely, the outskirts of M31 are dominated
by metal-rich stars ($<$[Fe/H]$>$ from $-$1 to $-$0.8, \citealt{Mould86,
Rich96, Durrell04}, and references therein).  It has been argued by
\citet{Kalirai06} that stars within 30 kpc of the center of M31 are
indeed part of an extended spheroid (or bulge). By itself, this property
illustrates the profound difference between M31 and the Milky Way. For the
Milky Way, such a chemically enriched extended spheroid which dominates
the star counts up to 30 kpc does not exist. In the following we review
briefly how the properties of the Milky Way outskirts (defined as a region
from 5 to 30 kpc from the centre) compare with those of external spirals.

Comparing with the properties of the outskirts of Milky Way and M31
however requires some care.  First, most abundance studies in the Milky
Way are based on high resolution spectroscopy while color magnitude
diagrams are most commonly used for stars in external galaxies. We
note however that the study of, e.g., \citet{Morrison03} shows that the
agreement between abundances estimated from spectroscopic and photometric
measurements are excellent. Second, the areas surveyed to derive stellar
abundances are small, they may be unrepresentative of the dominate
populations, and this may lead to biases in all estimates. For example,
an external observer of the Milky Way might have unluckily observed the
region where the bulk of the stars belonging to Sagittarius lie, and
would likely conclude that the outskirts of the Milky Way are indeed
enriched. However this alternative becomes more and more unrealistic
given the reasonable number and variety of locations of surveyed areas
in the outskirts of M31, and the insignificant fraction of the total
mass within Sagittarius.

\begin{figure}
\plotone{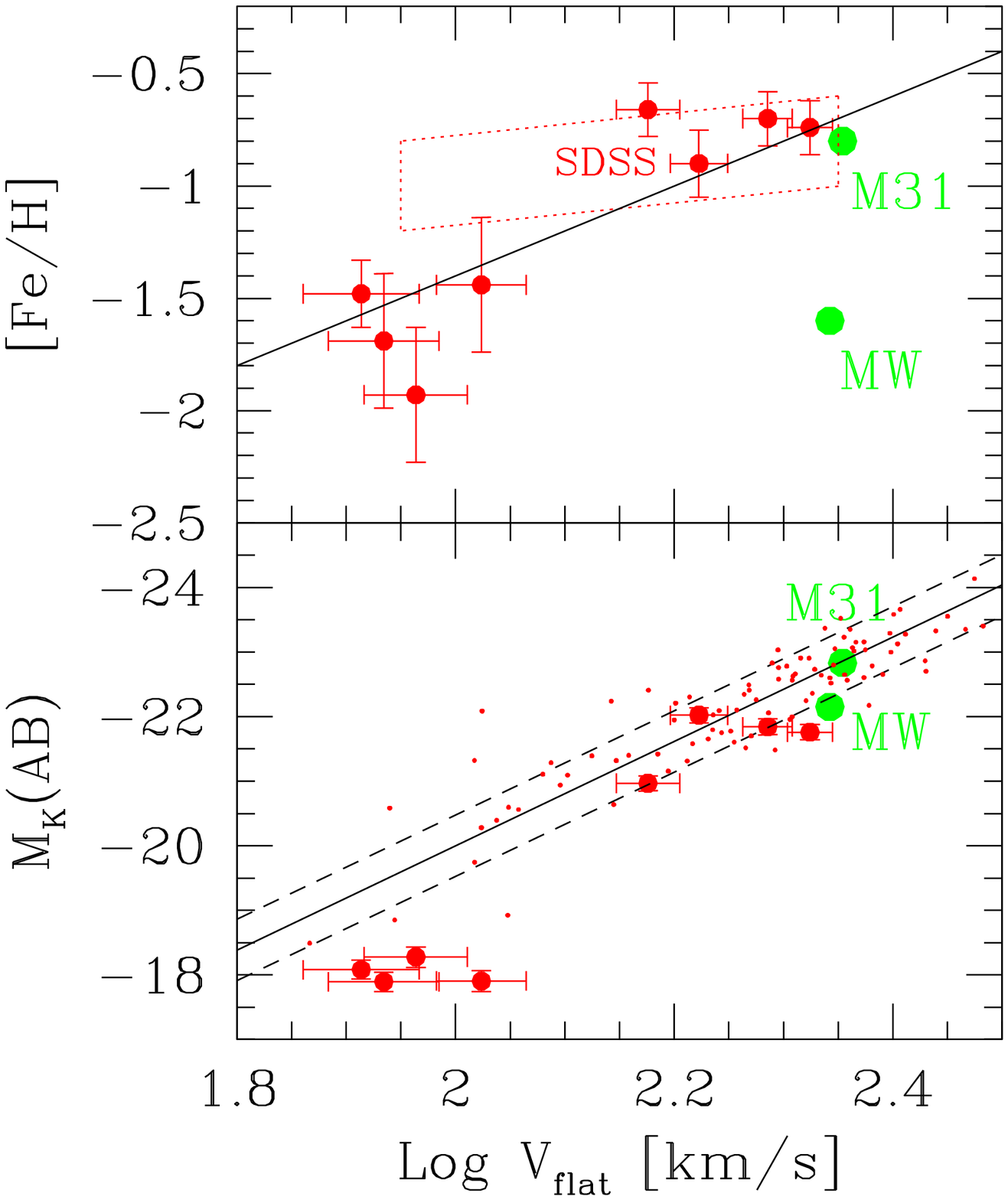}
\caption{{\it Top:}  Iron abundances estimated for the outskirts of
eight spiral galaxies from Mouhcine et al. (2006) plotted against
log($V_{flat}$) (small red points). Large green points represent the
values for the Milky Way and M31. The solid line assumes $M_{star}\sim$
$V^{4}$ \citep{McGaugh05} and $M_{star} \sim$  $Z^{2}$ following the
prescription of \citet{Dekel86}.  Red dashed lines represent the range
of \citet{Zibetti04}, after stacking 1047 edge-on SDSS galaxies,
and assuming that their colors are dominated by red giant stars.
{\it Bottom:} Tully-Fisher relation for the same galaxies \citep[very
small red dots represent the sample of][]{Pizagno06}.
\label{halo_FeH_TF2}}
\end{figure}

\citet{Mouhcine06} have compared the abundances of outskirts of several
spiral galaxies. \citeauthor{Mouhcine06} selected red-giant stars that
lie 2 to 10 kpc along the projected minor axis of eight nearby spirals,
in areas which are part of galactic outskirts although they might also be
contaminated by bulge or thick disk stars.  Fig.~\ref{halo_FeH_TF2} shows
the $<$[Fe/H]$>$ of red giant stars against $V_{flat}$:  all galaxies,
except the Milky Way, show a trend of an increasing metal abundance with
rotation velocity of the disk. This trend, found by Mouhcine et al.,
is likely explained through an examination of Fig.~\ref{halo_FeH_TF2}
(bottom): while massive galaxies have transformed most of their gas into
stars and metals, smaller galaxies have been much less efficient in doing
so, and still include a large gas fraction as indicated by their location
in the Tully-Fisher relation \citep[e.g.,][]{McGaugh05}. Interestingly
such a relation, if confirmed, requires a certain mixing of stars with
different enrichment patterns in the outskirts of the galaxy, such as
might be provided by a merger. The Galaxy's outskirts are under-abundant
relative to the trend line formed by external galaxies by about 1 dex
implying that it has been far less enriched than those of other galaxies
of the same total mass. On the other hand, M31 shares a similar location
than other large spirals in Fig.~\ref{halo_FeH_TF2}. This strengthens
our hypothesis that the properties of M31 are rather typical of large
spiral galaxies while the Milky Way appears to be exceptional.

Similarly, and perhaps more generally, the results of \citet{Zibetti04}
add more credence to our hypothesis. They found, by stacking 1047 images
of SDSS edge-on spirals, that the average color of stars at $\ge$ 5
times the disk scale length, beyond the disk minor axis, is redder by
$\Delta$(r-i)=0.3-0.4 than Milky Way globular clusters (or Galactic
halo stars), or after converting SDSS photometry \citep{Jordi06},
by $\Delta$(V-I)= 0.45$\pm$0.1. This discrepancy is especially large
for the brightest galaxies of the \citeauthor{Zibetti04} sample, i.e.,
those with similar absolute magnitudes as the Milky Way. The brightest
galaxies show the reddest stellar halo colors, accentuating this
difference with the Milky Way. \citeauthor{Zibetti04} show convincingly
that their measurements are not significantly affected by dust. If these
measurements are dominated by red giant stars, such a large shift in (V-I)
colors is unlikely related to an age difference between SDSS galaxies
and the Milky Way,  but more likely due to different metallicities (see,
e.g., Lee et al. 1993). The Galactic globular clusters and halo stars
are indeed very old, and a 0.45 mag shift in V-I to the red is likely
due to a shift of 0.8 dex$\pm$0.2 in [Fe/H] \citep{Lee93}. Comparing the
(V-I) colors from Zibetti et al. to that of nearby galaxies studied
by Mouhcine et al. allows us to place this ensemble of galaxies in
Fig.~\ref{halo_FeH_TF2}.  It is apparent that the SDSS galaxies lie within
the range defined by the intermediate mass spirals, including M31. The
bulk of the Zibetti et al. (2004) sample is composed of galaxies with
$M_{i}$ (where i refers to the i band of the SDSS) ranging from $-$19.5
to $-$22.5 (for $H_0$=70 km s$^{-1}$ Mpc$^{-1}$).  This corresponds
to log($V_{flat}$) ranging from 1.95 to 2.35 or galaxies with rotation
velocities similar to less than that of M31 and the MW \citep{Pizagno05}.

Combined together, the above results strongly suggest that stars
in the outskirts of the Milky Way have an average chemical abundance
three times lower than those of most spirals within a similar mass
range. Simulations show that the very low metal abundance of the outskirts
of our galaxy may require the absence of any previous merger of satellites
with mass larger or equal to $10^{9}$ $M_{\odot}$ \citep{Font06}.

\section{Towards a formation scheme for spiral galaxies within the
context of differences between the Milky Way and M31}

In the previous sections of this paper, we have shown that perhaps
the properties of the Milky Way are not representative of those of a
typical local spiral galaxy.  Within this context, we have used M31
as a foil to the Milky Way trying to demonstrate that it is a typical
representative of the spiral galaxy population (at least for those
that have rotation speeds similar to the Milky Way).  Since we find a
compelling case for the differing relative natures of the Milky Way and
M31, we wish now to discuss and conjecture on what these differences
may be telling us about galaxy formation.  Unlike previous discussions
of the relative characteristics of the Milky Way and M31, we wish to
embed this within the context of what we know about the properties of
distant galaxies.

\subsection{A quiet formation history compared to a merger-driven
formation history: the MW versus M31}

Historically, the Milky Way and M31 were thought to be quite
similar. They have the same Hubble type and similar total
masses. At a given $V_{flat}$, we find that M31 has a stellar mass and
angular momentum close to that of the average of local spiral galaxies. In
contrast with this, the Milky Way has a stellar mass and angular
momentum that are two to three times lower than average. Its
outskirts could be even more peculiar, with stellar abundances three
times lower than those of other spirals with approximately the same
absolute magnitude and rotation velocity.

What renders the Milky Way so exceptional? Its peculiarities seem to be
tightly linked to its ``quiet'' history of formation. It is interesting to
compare its formation history with those of other local spirals galaxies. M31
has properties that are much closer to those of an average spiral. Let
us investigate how M31, when compared to the Milky Way, may have acquired
2 times more stellar mass and 2.5 times more angular momentum.

Detailed studies of bulge, disk and outskirts of M31 have been (mainly
recently) made.  Recent discoveries about M31 are numerous (giant
stream, large faint clumpy disk, age-metallicity of stars in the disk,
characterization of the metallicities and ages of the globular cluster
system, detailed measurements of the chemical abundances and ages of
bulge and halo stars, etc.).  Reviewing these discoveries alone would
by itself require a very long manuscript.  Such a detailed review
is obviously beyond the scope of this paper. Let us just summarise some
of the widely accepted implications of these new discoveries. Both the
stream \citep{Ibata01} and the extended clumpy thick disk \citep{Ibata05}
plead for an active merger history for M31. \citet{Ibata05} suggested
that either a succession of minor mergers or possibly a major merger,
in both cases having occurred within the last 8 Gyrs, could explain these
properties of M31.

Studies of color magnitude diagrams of disk, bulge, and halo
stars provide a complementary view of the star formation history
of M31.  The ages for globular clusters derived from Lick indices
are sensitive to the template model used for the analysis
\citep[see, e.g.,][]{Beasley04}. For example, using \citet{Thomas03} or
\citet{Bruzual03} templates change the ages of the intermediate ages of
globular clusters discovered by \citet{Beasley04}, from $\sim$ 3 to $\sim$
6 Gyrs. Nevertheless it is widely accepted that the globular cluster
system of M31 resembles that of the Milky Way, on which is
superposed a population of globular clusters with intermediate ages
(from less than 1 Gyr to about 8 Gyr). Interestingly, such globular
clusters with intermediate ages can be formed during advanced, but still
relatively obvious mergers \citep[e.g.,][]{Schweizer06}. In M31 this
additional population represents $\sim$42\% of the whole GC systems and
reflects a relatively recent enrichment of the outskirts of M31 by infall
of another galaxy \citep{Beasley04}. This is confirmed by the large age
variance found for the properties of the stars in the outskirts \citep[or
``spheroid'', see][]{Brown06} of M31, which contrasts with the single, old
stellar population in the stars of our own galaxy's outskirts. The
stream star ages show a peak at $\sim$8-9 Gyrs \citep{Brown06}, while
the (outer) disk stars show a peak at 5-7 Gyrs, about 2 Gyrs younger.

The differences between the Milky Way and M31 are likely due to the
quiescent formation in the former case and to the merger dominated history
for the latter. It is still not definite whether M31 has experienced
a major merger or a succession of several minor mergers. Both merger
scenarios can explain the significant fraction of stars with intermediate
age and metallicity in the outskirts, as well as the fact that M31 has
assembled twice the stellar mass of the Milky Way. For either alternatives,
successive minor mergers (or many episodes of intense minor merging) or a
few major mergers over the lifetime of a galaxy, simulations reproducing
all the above observations are however not yet available.
Besides this, the large angular momentum of M31 compared to the Milky
Way (greater by a factor of 2.5), suggests a major encounter.
Successive and numerous minor mergers are likely to be less efficient in
producing large angular momentum\footnote{Note that this argument also
applies if one considers the spin parameter, because the total masses
of M31 and Milky Way being so similar, their ratio of spin parameters
scales with that of their angular momentum.}.  Another argument, given
by \citet{Brown06}, is: ``if the Andromeda spheroid is the result of
many smaller mergers that did not occur in the Milky Way, one must
ask why there is such a statistically significant distinction between
the merger histories of two similarly-sized spirals in the same galaxy
group.''  \citet{Ibata05} also favored a major encounter to interpret
the homogeneous chemical properties of the ``extended clumpy disk''.

How credible is the hypothesis that M31 had experienced a major
merger in the past? If the M31 stream is a relic of such an event,
it might have occurred $\sim$ 8 Gyrs ago, as was suggested by
\citet{Brown03}. Simulations of the stream have focused on a
very recent (0.5 Gyr old) merging of a $10^{9}$ $M_{\odot}$ dwarf
\citep{Font06}. \citet{Brown06} suggested it would be useful to test
whether the stream could actually be due an earlier event in the
history of M31, as well as whether such a giant stream would still
be evident after 8 Gyrs. Indeed, such a major merger can explain the
similarities between the stream and the populations of the outskirts
\citep{Brown06} since both will be polluted in a similar way during the
merging process. This is in contrast with expectations from a scenario
involving multiple minor mergers for which age-metal signature should be
far less uniform that what it is observed. Eight Gyrs ago, the progenitor
of M31 was unavoidably a much smaller galaxy than it is today: as noticed
by \citet{Ibata05}, an event involving a galaxy of the M33 size would have
been a major merger event. Is such an ancient major merger event realistic
knowing that M31 possesses a large thin disk \citep{Morrison04}? On the
basis of the properties of the old disk globular clusters in comparison,
\citet{Morrison04} have argued that the thin disk is indeed old. However,
their arguments have been contradicted by much lower age estimates of
the same objects \citep{Beasley04}.  The ages of disk stars, 5-7 Gyrs,
are consistent with a disk built predominately during $\le$ 2 Gyrs after
a major merger event: simulations predict a rapid formation of the new
disk after a major gas-rich merger event \citep[but also including the
effects of substantial feedback][]{Governato06, Robertson06}.

The large (specific) angular momentum of M31 (compared to that of the
Milky Way), the similarity of halo and stream stellar populations, the
difficulty in having a succession of many minor encounters in M31 and
very few or none for the Milky Way, all  favour a relatively recent,
$\sim$8 Gyrs ago, major merger having a substantial impact on the final
characteristics of M31.  Interestingly a similar conclusion has been
reached by \citet{Durrell04}. They noticed that a major merger scenario
naturally explains the relative metallicity distribution functions
of both the stars in the outskirts and globular clusters. They also
suggest similarities between the M31 spheroid and the outer part
of giant ellipticals. As a matter of interest, M31 falls precisely
on the relation between black hole and bulge mass as for elliptical
galaxies \citep[e.g.,][]{Tremaine02}. This suggests that, even on much
smaller scales than that of the outskirts, a major merger might have
had a substantial impact on the properties of M31.  Notice specifically
that the black hole-bulge relationship has been explained in the frame
of major mergers, intimately connecting a quasar phase of galaxies
\citep{Hopkins06a, Hopkins06b} to the overall evolution of galaxies
\citep{Springel05} with AGN feedback being a necessary and crucial
process responsible for this interplay.

In summary, one can explain the differences between the Milky Way and M31
by the absence of significant merger event(s) over the last 10-11 Gyrs for
the former. M31 has a stellar mass and a chemical enrichment in the stars
in its outskirts typical of local spiral galaxies. Its angular momentum
is 1.25 times that of an average SDSS galaxy at same $V_{flat}$. Besides
this, the Milky Way has a deficiency of a factor two in both stellar mass
and angular momentum, when compared to a similar average. Accounting for
this, galaxies as or more exceptional than the Milky Way represent only
7\% of all local spirals. Because of the remarkable properties of the
relatively unpolluted Galactic outskirts, the fraction of Milky Way-like
galaxies may be even lower. It is then probable that most local spirals,
like M31, have experienced more and perhaps, later mergers than the
Milky Way. In such a case, the differences in the formation histories
of the Milky Way and M31 are simply reflected in the differences between
the properties of the Milky Way and those of the bulk of local spirals.

\subsection{Does the Exceptionally Quiet Merger History of the MW Imply
it Grew through Secular Processes?}

Advocating a new scenario of spiral formation means we have to break
some ``taboos'' -- the main one being the fact that Milky Way may be an
exception rather than the rule. The widely accepted assumption that
a major merger would unavoidably lead to an elliptical is perhaps no longer
tenable: accounting for the large number of major mergers that have
apparently occurred since z=3 would imply that all present-day galaxies
should be ellipticals.  This is obviously not the case. So it is likely
that disks either can survive or are ``rebuilt'' after a major merger,
through whatever mechanism as yet perhaps unknown in detail \citep[see,
for example,][]{Robertson06}.  

Many simulations have the implicit assumption that most of the star
formation in intermediate mass galaxies had occurred well before z=1. This
is however not correct: nearly half of their present-day stellar masses
has been formed since that epoch, as shown by \citet{Hammer05}, or
\citet{Bell05}. But more than this, to reproduce the zero-point of the
local Tully-Fisher relation within the framework of the standard gas
accretion model, requires spin parameters ($\lambda$ from 0.06 to 0.08)
larger than those expected for dark matter halos \citep[$\lambda$=
0.042 on average, see][]{Pizagno05, Dutton06}. This discrepancy has
led \citet{Dutton06} to relax the adiabatic contraction hypothesis of
the standard scenario. They assume instead that the halo expands during
a major merger and must lead to a new disk being formed (or perhaps a
pre-existing disk preserved in some way), with a clear reference to the
simulations of \citet{Robertson06}.  It seems more and more evident that
the formation of disk galaxies requires a larger influence of mergers
than hypothesized in the standard scenario.

The Milky Way being an exceptionally quiet galaxy alters the validity of
the standard scenario of spiral formation which has been mostly based on
our Galaxy's properties. This being the case, the situation is in reality
worse for the general validity of this scenario.  Indeed, giant stars
of the Galactic bulge show large $\alpha$/Fe ratio indicating a fast
($\sim$1 Gyr) formation of the bulge \citep{Zoccali06, Lecureur06}. If
confirmed, it seems that the Milky Way's bulge formed at very early epochs
through the merger of large clumps (or progenitor galaxies), with the
disk being built (or rebuilt) later on. This is in stark contrast with
a primordial condensation of gas into a disk which then forms stars,
the main assumption in the standard scenario.

More interesting perhaps is to consider the Milky Way as an archetype
of a galaxy having experienced no merging event since the last 10-11
Gyrs. We assume here that very small encounters such as the present
disruption of Sagittarius (with a mass of less than 1\% of that of the
Milky Way) are not sufficient to significantly enrich the stellar halo
in mass or substantially alter its average metallicity. \citet{Font06}
indeed  argue that the very low metal abundance of the Galactic stellar
halo requires the absence of any previous merger of satellites with mass
larger or equal to $10^{9}$ $M_{\odot}$. If correct, this implies that
the Galactic disk may have mostly grown by an approximately smooth gas
accretion, or, in other words, by secular evolution \citep[see also]
[who suggest that such infall may be sufficient to build the Milky
Way disk]{Croton06}.  In the following, we consider that secular evolution 
includes either smooth gas accretion or accretion of small satellites.
 Let us consider that the Galactic disk has been
formed by secular evolution and that the Galactic bulge \citep[in which
$\sim$ 25\% of the stellar mass is locked, see][]{Flynn06} was formed by
an early merger. If the fraction of the mass of both the Milky Way and
M31 that grew through secular evolution -- accretion of gas and of small
satellites -- have been similar, since they inhabit the same group, we can
use the difference in stellar masses to estimate the likely contribution
of smooth gas accretion in general. The difference between M31 and Milky
Way stellar masses can be accounted if $\sim$36\% (estimated with the
ratio $M_{star}^{MW disk}$/$M_{star}^{M31}$) of the mass assembly of
M31 is due to secular evolution, while 64\% may be directly linked with
mergers. Since M31 has a typical stellar mass amongst local spirals
of the same total mass (also including the dark matter component which
influences their dynamics), this balance may apply to most spirals, or at
least those inhabiting environments of similar density as the Local Group.

\subsection{The Spiral Rebuilding Hypothesis: Formation of spiral galaxies
after major mergers?}

The mass assembly of typical spirals, including M31, has been probably
driven predominantly by mergers and that their assembly history might not be
best represented in the characteristics of the Milky Way. Galaxies with
approximately the same rotation speeds as the MW also show larger angular
momentum. Those can be produced by a single major merger while it is difficult
to reconcile with a succession of minor mergers. Here we investigate
how a scenario with major mergers can be reconciled with observations.

The spiral rebuilding scenario was proposed by \citet{Hammer05} to explain
the observations of the distant galaxies. Specifically, this hypothesis
was used to explain for distant galaxies the simultaneous evolution
of the global stellar mass, Luminosity-Metallicity relationship, pair
statistics, evolution of the IR light density, colors of spiral cores
(bulges?), evolution in the number density of spheroids and spiral
galaxies, and evolution in the fraction of peculiar galaxies (mergers
and compact). The spiral rebuilding scenario is consistent with all these evolutionary trends,
while a scenario for which the stellar mass formation is dominated by
minor encounters \citep[``collisional starbursts'',][]{Somerville01},
has difficulties in reproducing in particular the evolution of the IR
light density, number density of peculiar galaxies and spiral core colors.

In such a framework, the question of the representativeness of the Milky
Way may simply be explained by the low fraction of galaxies that have
escaped a major merger since z$\sim$3.
 Galaxies like the MW with a quiet merger history are expected
to have, on average, a lower stellar mass, a lower angular momentum and
a less enriched stellar halo for their rotation speed, but still could be
the product of a major merger. Conversely, the representativeness of M31
compared to the 50 to 75\% fraction of galaxies of similar masses, may be
a result of the ubiquity of major mergers since z=1 (see \S 1) and their
significant influence in determining the properties of spirals at the
current epoch.  The differences between the MW and M31 may be therefore
simply due to the epoch of the last equal or nearly equal mass merger.
In the case of the MW, perhaps the last major merger occurred well before
z=1, while for M31 it is likely that it occurred around or after z=1.

Major advances in simulations have provided a theoretical framework for
the disk rebuilding scenario. \citet{Cox06} have shown that the remains
of dissipational mergers have significant rotation and angular momentum
compared to dissipation-less mergers. Even without efficient feedback,
ellipticals, when formed after a merger, possess a seed for the subsequent
formation of a rotationally supported disk. \citet{Robertson06} have shown
that gas dominated mergers (gas fraction larger than 50\%) can produce
remnants with disks of sufficiently high angular momentum \citep[unlike
simulations that do not include gas-rich major mergers to explain the
large angular momentum of disks; e.g.,][] {Steinmetz99}. The importance
of high gas fraction has been already been suggested by \citet{Hammer05},
since it is the gas expelled through the impact of strong feedback which
subsequently feeds the newly formed disk and provides the necessary
rotational support and sufficient specific angular momentum \citep[e.g.,
][]{Robertson06}. Since stellar mass density of galaxies has nearly
doubled since z=1, the condition that galaxies have large gas fractions
must be the case on average. Evolution of the gas content in galaxies is
also observed, although indirectly, from the observed evolution of the gas
phase metal abundance in distant galaxies. \citet{Liang06} have estimated
that the gas content in intermediate mass galaxies at z$\sim$0.6 was
two times larger than in galaxies at the current epoch. \citet{Erb06}
have found, at z$\sim$ 2, gas fractions ranging from 0.2 to 0.8.

Although the consequences of adopting this scenario need to be more accurately evaluated, the
rebuilding disk scenario is becoming more and more viable. Regardless,
observations show that there is sufficient gas at redshifts
less than one out of which disks could be rebuilt.  The rebuilt disks
and bulges after a major merger should be compared to observations of
present-day galaxies in a realistic way, i.e., after accounting for
further gas accretion (or secular evolution) which may essentially feed
the disk \citep[see also][]{Croton06}.  Thus the rebuilding disk scenario
may only require some tuning of the assumptions already made in most
simulations. In essence,
this scenario implicitly solves both problems of the standard scenario
(e.g., disk stability to further collisions and angular momentum),
because subsequent collisions help to generate large angular momentum as
it is observed in local spirals.  Late epoch merging generally results
in disks with higher angular momentum than disk which had their last
major merger earlier in the history of the Universe as perhaps did the
Milky Way.  Besides explaining the difference it angular momentum, it
may also explain why the average metallicity of stars in our halo are
less than that of typical spirals at the same rotation speed and other
characteristics of the Milky Way's halo.

However, this alternative to the standard scenario has not been carefully
considered in the literature or through simulations, possibly because
it appears too exotic or too disturbing. A possible caveat could be the
significant fraction of LIRGs (galaxies with high IR luminosities of $\ga$
10$^{11}$ L$_{\sun}$) at high redshift showing spiral morphologies
\citep{Melbourne05, Bell05}. However, the rebuilding disk phase
corresponds to a phase of strong gas infall (from gas left over from the
merger and by a gradual cooling of the hot halo gas), during which the
galaxy may well have the appearance of a LIRGs with spiral morphology. 

Is there observational evidence for a subsequent disk rebuilding phase
occurring later, after the merging? At least an interesting clue can be
derived from recent observations of the evolution of the Tully-Fisher
relation. One can compare the pair fraction (5$\pm$1\% of z=0.6 galaxies,
see \S 1) to the higher fraction \citep[26\%,][]{Flores06} of z=0.6
galaxies having complex velocity fields, the later being probably
associated with merger remnants from comparison to numerical simulations
\citep{Puech07}. Assuming 0.35 Gyr as the characteristic time for a
real pair to actually merge (see \S 1), implies a remnant phase lasting
from 1.5 to 2 Gyrs, in a relatively good agreement with expected times
needed to rebuild a significant disk after an efficient feedback phase
\citep{Robertson06, Governato06}. A more quiescent history of cold
and clumpy gas flow has been also been investigated for disk formation
\citep{Dekel06}. Such a scenario however needs to show how it succeeds
in reproducing the strong evolution of violent star formation with epoch
(i.e., the strong number density evolution in galaxies with substantial
IR luminosities, L$_{IR}$$\ga$ 10$^{11}$ L$_{\sun}$) and in solving the
angular momentum problem. Assuming collisions of large enough gas clumps
might succeed in explaining the the characteristics of distant galaxies,
although it would at any rate be close to a scenario hypothesizing a
large fraction of major mergers at intermediate redshifts as is observed.

\section{Conclusion}

Compared to local spirals with similar rotation velocities, the Milky
Way has a significant deficiency of stellar mass, angular momentum and
chemical abundances in stars in its outskirts. We have argued that
these differences can be
interpreted as being due to the exceptionally quiet formation history
of the Milky Way compared to other spirals at comparable rotation
velocities. After the rapid building of its bulge, more than 10 Gyr ago,
the disk of the Milky Way has been formed inside-out mostly through smooth
gas accretion, in a secular mode. In the same dynamical mass range, the
bulk of spiral galaxies, including M31, have accreted most of their mass
and angular momentum through a more recent and active merger history. In
other words, other galaxies may have populations of stars in their
outskirts similar to the Milky Way but to which a significant component
would have been accreted later through further episodes of mergers.
Combining results from observations of distant galaxies (merger rate
and evolution of LIRGs) to those of local galaxies (Tully-Fisher and
angular momentum), we have hypothesized that most spiral galaxy disks
are ``rebuilt''.  The timing of this rebuilding and how many episodes of
rebuilding are also crucial parameters within this hypothesis.  In such
a scenario, to reproduce the properties of the Milky Way, the disk of
the Milky Way was built (or rebuilt if it had a pre-existing disk) at a
much earlier epoch ($\sim$10 Gyrs ago) than the general population of
local spiral galaxies.  Thus, the standard scenario of disk formation
may not even apply to the Milky Way.

The major advantages of the disk rebuilding scenario are that it may solve
simultaneously various problems such as the so-called angular momentum
problem (major merger are very efficient to produce angular momentum),
the large fraction of stellar mass produced in LIRGs since z=1 (major
merger are an efficient way to produce episodic strong star formation),
and the chemical abundances of stars in the outskirts of spiral galaxies
(by efficient mixing).  However, the most important point of this paper
is that it may also naturally explain why the Milky Way is so exceptional
in its properties when compared to other spirals. Specifically, we
think that adopting the spiral rebuilding scenario for understanding the
evolution of spiral has a number of significant consequences.  These are:

\begin{itemize}

\item M31 appears to be a typical spiral, emphasising the need to
investigate further its precise formation history in order to understand
how the majority of spiral galaxies may have formed;  What differences
in the final characteristics result from the exact time at which the 
rebuilding occurs?

\item That the Milky Way had an exceptionally quiet formation history,
having escaped any major merger (and possibly a significant number
of minor mergers) during the last 10-11 Gyrs; Milky Way like galaxies
correspond to only 7\% of local spirals, and possibly much less;

\item Suggests the need for modelling the formation history of spirals
using a superposition of a ``quiet'' history like that of the Milky
Way with an active history of merging, which could be responsible for
$\sim$2/3 (perhaps more) of the stellar mass;

\item Explains the failure of the standard scenario for spiral formation
which is unable to reproduce the properties of spiral galaxies,
including their stellar mass-Tully-Fisher relation, their high angular
momentum and the chemical abundances of stars in their outskirts;

\item There is a reasonable chance that the spiral rebuilding scenario can
reproduce most of the observed properties of local spirals, as well as
those of their progenitors at large distances.

\end{itemize}

Galaxy simulations are certainly needed to investigate the accretion history of
spiral galaxies. The evidence for a scenario where the evolution of
disk galaxies is driven predominantly by mergers rather than by other
secular processes, allows us to suggest that the key question is to now
gauge the relative impact of minor and major merger in shaping spiral
galaxies as we seen them today. Observationally, M31 is certainly the
best target for a robust investigation to determine the past history
of a typical spiral galaxy. In addition, follow-up of the pioneering
modelling work of \citet{Robertson06} and \citet{Governato06}
would be also very valuable in understanding the role of mergers (both
major and minor) in disk galaxy evolution. Major mergers with a variety
of mass ratios and angular momentum vectors should be investigated to see
if then can reproduce the characteristics of spiral galaxies in detail.
Such simulations must include realistic amounts of gas for the progenitors
within a redshift range z=0.5-3 and more observational effort should be
expended in determining this parameter. Subsequently, such models could
be scaled to the observed fraction of mergers (and merger remnants) and
by observationally determined distributions of mass ratios and angular
momentum vectors. By assuming after the last major merger, a realistic
rate of ``smooth'' gas accretion, would let one investigate whether
the re-shaped disks and bulges are consistent (or not) with the observed
properties of the ensemble of bulges and disks in present-day galaxies.

\acknowledgments
The authors wish to thank F. Spite and M. Spite, and especially R. Ibata
and A. Robin, for helpful comments, discussions, and carefully reading
several versions of this manuscript.  Their effort is greatly appreciated.
We thank M. Steinmetz for kindly sharing the results of his work with
us.

\appendix

\section{On the Rotational Velocity of the Milky Way}

As noticed by Flynn et al (2006), the Tully Fisher relation is so steep,
that with a value of $V_{flat}$=185 km s$^{-1}$, the Milky Way would
nicely fall along and within the scatter of the Tully-Fisher relation of
external galaxies (see Fig.~\ref{TF_all}). Indeed estimates of the local
circular rotation, $\Theta_{0}$, values down to 185 km s$^{-1}$ (Olling \&
Merrifield, 1998) and up to 235 km s$^{-1}$ (Reid \& Brunthaler, 2004)
or even 255 km s$^{-1}$ (Uemura et al., 2000) have been reported. These
estimates are highly degenerate as they depend on the value of $R_{0}$ --
the distance to the Galactic Center.  This is well illustrated in Figure~1
of Olling \& Merrifield (1998). Here we consider various combinations
of ($\Theta_{0}$, $R_{0}$) determined or adopted by different studies. They
are (185, 7.1), (235, 8.0) and (255, 8.5) for Olling \& Merrifield (1988),
Reid \& Brunthaler (2004) and Uemura et al. (2000), respectively. The most
accurate and direct estimate of $R_{0}$(= 7.94$\pm$0.42 kpc) has been
determined by Eisenhauer et al. (2003) on the basis of proper motions
around the black hole in the Galactic center.  This
estimate was further refined
in \citet[][$R_{0}$= 7.62$\pm$0.32 kpc]{Eisenhauer05}. Notice that
these values are in agreement with the best overall estimate of Reid
(1993) made by combinating previous estimates before the early
90s. Using the
Eisenhauer et al. (2005) value of  $R_{0}$, Brunthaler et al. (2006)
derive $\Theta_{0}$= 225$\pm$10 km s$^{-1}$, the uncertainty depending
mostly on the uncertainty in the estimates of $R_{0}$.

A robust measurement of the relative rotational velocity of the Sun to
the halo has been made by Sirko et al. (2004) who find 222.1$\pm$7.7 km
s$^{-1}$. This measurement was based on an analysis of the kinematics of
1170 blue horizontal branch stars in the Galactic halo. Solar velocities
significantly lower than the IAU standard would correspond to a halo
with significant counter rotation (Sirko et al. 2004). Conversely,
the rotation velocity of the Galactic halo is found to be in the same
sense as the disk rotation, and marginally consistent with 
no rotation (Sirko et al. 2004 and references therein). Extreme values for
the local circular velocity are also excluded by open cluster velocities
in the Milky Way (Frinchaboy \& Majewski 2005). The evidence seems to
favor a value of the rotation velocity of the Milky Way that is close
to the IAU standard and we see no need to adopt a different value.

To compare Milky Way to external galaxies requires us to determine what
an external observer would estimate for $V_{flat}$. Indeed within the
range of reasonable ($\Theta_{0}$, $R_{0}$) values (Brunthaler et al.
2006;  Eisenhauer et al. 2005), the fit of the Milky Way rotation curves
(see Fig.~1 of Olling \& Merrifield  1988) shows a flat  or a slightly
rising curve. To some extent, this contrasts with the rotation curve
of M31 which smoothly decrease from 259 km s$^{-1}$ at 10 kpc to 226 km
s$^{-1}$ at large radii (Carignan et al. 2006). Such a decline is likely
due to the effect of the prominent bulge of M31. Other considerations
are also very instructive. It has been argued
by Evans \& Wilkinson (2000) that, contrary to earlier ideas, the halo
mass of the Milky Way could be equal or even higher than that of M31.
Because $V_{flat}$ is intimately linked with the total mass of a galaxy
(see \S 3), under this hypothesis, it is reasonable to adopt for the Milky Way arotation speed (220 km
s$^{-1}$) that is close to that of M31 (226 km s$^{-1}$). In summary,
the deficiency of stellar mass of the Milky Way compared to M31, and
hence to external galaxies, appears to be particularly robust.

\end{document}